\documentclass[a4paper,12pt]{article}
\usepackage{amsmath,amssymb,amsthm,epic,eepic,graphicx}
\usepackage[english]{babel}
\usepackage{hyperref}
\title{Dressing chain for the acoustic spectral problem}
\author{V.E. Adler, A.B. Shabat}
\date{\normalsize
 L.D.~Landau Institute for Theoretical Physics, RAS\\
 1-A pr.~Ak.~Semenova, 142432 Chernogolovka, Russia \\
 E-mail: {\tt adler@itp.ac.ru}}
\advance\textwidth 20mm  \advance\oddsidemargin -10mm
\advance\textheight 20mm \advance\topmargin -15mm

\def\a{\alpha}
\def\b{\beta}
\def\g{\gamma}
\def\d{\delta} \def\D{\Delta}
\def\eps{\varepsilon}
\def\phi{\varphi}
\def\ka{\varkappa}
\def\la{\lambda}
\def\pd{\partial}
\def\ti{\widetilde}

\def\<{\langle}\def\>{\rangle}
\def\const{\mathop{\rm const}}
\def\sech{\mathop{\rm sech}}

\newtheorem{statement}{Statement}
\newtheorem{rem}{Remark}
\newenvironment{remark}{\begin{rem}\rm}{\end{rem}}

\begin{document}\maketitle\thispagestyle{empty}
\begin{abstract}
The iterations are studied of the Darboux transformation for the generalized
Schr\"odinger operator. The applications to the Dym and Camassa-Holm
equations are considered.
\end{abstract}

\section{Introduction}

In this paper we consider the generalized Schr\"odinger problem
\[
 \phi_{yy}=(q(y)-\la r^4(y))\phi
\]
and associated Dym \cite{Kruskal} and Camassa-Holm equations
\cite{Fokas_Fuchssteiner,Camassa_Holm}. Although this problem can be brought
into the standard form $r=1$ via the Liouville transformation, the
reformulating of the known results is not a quite trivial job, due to the
inevitable change of independent variable. This results, in particular, in
the fact that all known exact solutions cannot be given explicitly and are
represented in parametric form. The scattering theory was developed, among
the others, in the papers \cite{Krein52,BSS98,Constantin01}. The papers
\cite{Dmitrieva93a,Dmitrieva93b,Constantin_McKean99,Novikov,Alber_Fedorov01,ACFHM}
and others were devoted to the construction of algebraic-geometric,
multisoliton and peakon solutions. The Darboux-B\"acklund transformations
were considered in the papers \cite{Schiff,Hone,Ivanov}. However, from our
point of view, the techniques of these transformations remains so far
inadequate. Meanwhile, the Darboux transformations for the Schr\"odinger
operator have proven themselves as an effective tool for constructing of
exactly solvable potentials and explicit solutions of the KdV equation, see
eg. \cite{Crum,Lamb80,Shabat92,Veselov_Shabat} and many other works. It is
not always possible to copy the results obtained here, because of the actual
differences in the settings of the spectral problems, but the main ideas
should be reproduced. This is what we try to do in this article.

The presence of two potentials allows to combine the Darboux and Liouville
transformations. This possibility yields interesting consequences and
deserves detailed study. The key observation of the presented paper is that
one of such combinations is almost as simple as the Darboux transformation
itself but corresponds to the gauge $q=0$ rather than $r=1$.

\section{Darboux transformations}\label{sec:DT}

The goal of this Section is to bring into the consideration the transforms
acting on the set of the generalized Scr\"odinger equations
\begin{equation}\label{Dy:rq}
 \phi_{yy}=(q(y)-\la r^4(y))\phi.
\end{equation}
Let us start from the classical transformation which eliminates the factor
$r$. We reserve the own notation for this form of equation, because of its
importance.

\begin{statement}[Liouville transformation]\label{st:LT}
Equation (\ref{Dy:rq}) takes the form
\begin{equation}\label{Dx:u}
 \psi_{xx}=(u(x)-\la)\psi
\end{equation}
under the change
\begin{equation}\label{LT}
 dx=r^2dy,\quad \psi=r\phi,\quad u=q/r^4+r_{xx}/r.
\end{equation}
\end{statement}

Another classical transformation is defined by any particular solution
$\psi^{(\a)}$ of equation (\ref{Dx:u}) at $\la=\a$. It is not difficult to
give it in the general gauge (\ref{Dy:rq}) as well.

\begin{statement}[Darboux transformation]\label{st:DT1}
Equation (\ref{Dx:u}) is form invariant under the transformation
\begin{equation}\label{DT1:uf}
 \hat\psi=\psi_x-f\psi,\quad f:=\psi^{(\a)}_x/\psi^{(\a)},\quad
 f_x+f^2=u-\a,\quad \hat u=u-2f_x.
\end{equation}
Equation (\ref{Dy:rq}) is form invariant under the transformation
\begin{equation*}\label{DT1:rq}
\begin{gathered}
 \hat\phi=\frac1{r^2}(\phi_y-f\phi),\quad f:=\phi^{(\a)}_y/\phi^{(\a)},\quad
 f_y+f^2=q-\a r^4,\\
 \hat r=r,\quad
 \hat q=q-2f_y+\frac{4r_y}{r}f+6\frac{r^2_y}{r^2}-2\frac{r_{yy}}{r}.
\end{gathered}
\end{equation*}
\end{statement}

Now let us consider the other canonical form of the equation (\ref{Dy:rq}),
\begin{equation}\label{Dy:r}
 \phi_{yy}=-\la r^4(y)\phi.
\end{equation}
The so called acoustic spectral problem corresponds to the potential $r$
bounded away from $0$ and tending to $1$ rapidly enough at $|y|\to\infty$,
\cite{BSS98}. However, we will not care of the analytic properties at first
and we will refer to (\ref{Dy:r}) just as to acoustic equation. Obviously,
it is obtained from (\ref{Dx:u}) by inverse Liouville transformation, if one
choose $q=0$ and take a wave function at $\la=0$ as $r$:
\begin{equation}\label{LT0}
 dx=r^2dy,\quad \psi=r\phi,\quad u=r_{xx}/r.
\end{equation}
It is easy to show that the arbitrariness in the choice of $r$ results in
the linear-fractional changes on the set of equations (\ref{Dy:r}):
\begin{equation}\label{Mobius}
 \ti y=\frac{c_1y+c_2}{c_3y+c_4},\quad \ti\phi=\frac{1}{c_3y+c_4}\phi,\quad
 \ti r=\frac{c_3y+c_4}{\D^{1/2}}r,\quad \D=c_1c_4-c_2c_3.
\end{equation}
The recomputation of the Darboux transform brings to unexpectedly simple
formulae.

\begin{statement}\label{st:DT2}
Equation (\ref{Dy:r}) is form invariant under the transformation
\begin{equation}\label{DT2}
\begin{gathered}
 \hat\phi=\phi_y/p-\phi,\quad p:=\phi^{(\a)}_y/\phi^{(\a)},\quad
 p_y+p^2=-\a r^4,\\
 \hat r=p/r,\quad \hat r^2d\hat y=r^2dy.
\end{gathered}
\end{equation}
\end{statement}
\begin{proof}
Application of $D_{\hat y}=p^2r^{-4}D_y$ and elimination of $p_y$ in virtue
of Riccati equation yields first $\hat\phi_{\hat y}=\a\phi_y-\la p\phi$, and
next $\hat\phi_{\hat y\hat y}=-\la p^4r^{-4}\hat\phi$.
\end{proof}

In order to check that the presented transformation is equivalent to
(\ref{DT1:uf}), apply the change (\ref{LT0}) to $\phi$ and $\hat\phi$.
Taking into account the relation $f=\psi^{(\a)}_x/\psi^{(\a)}=r_x/r+p/r^2$
yields
\[
 \hat\psi =\hat r\hat\phi=\frac{p}{r}\Bigl(\frac1p\phi_y-\phi\Bigr)
 =r\phi_x-\frac{p}{r}\phi=r\Bigl(\frac{\psi}{r}\Bigr)_x-\frac{p\psi}{r^2}
 =\psi_x-f\psi.
\]

The equation (\ref{Dy:r}) is considered on a finite interval as well
\cite{Krein52}. It is possible to convert this case to the spectral problem
on the whole axis
\begin{equation}\label{Dz:R}
 \chi_{zz}=(1-\la R^4(z))\chi
\end{equation}
by means of the following Liouville transformation \cite{BSS98}:
\begin{equation}\label{LT:rR}
 y=\tanh z,\quad \phi=\chi\sech z,\quad r=R\cosh z.
\end{equation}
Here we assume that $r\in C^\infty([-1,1])$, $r>0$, $r(-1)=r(1)$. The
importance of this gauge is caused by the relation to Camassa-Holm equation
which is discussed in \hyperref[sec:CH]{Section~\ref*{sec:CH}}. However, the
Darboux transformation looks rather awkward in these variables, and it is
more convenient to keep using the formulae (\ref{DT2}), recalculating the
answer by the indicated substitution.

In conclusion, we mention two more simple auto-transformations of the
acoustic equation.

\begin{statement}\label{st:invshift}
The form of equation (\ref{Dy:r}) does not change under the
transformations
\begin{gather}
\label{inv}
 \bar\phi=\phi_y,\quad \bar r=1/r,\quad d\bar y=r^4dy,\\
\nonumber
 \ti\phi=\phi/\phi^{(\a)},\quad \ti r=r\phi^{(\a)},\quad
 (\phi^{(\a)})^2d\ti y=dy,\quad \ti\la=\la-\a.
\end{gather}
\end{statement}
\begin{proof}
One has for the first transformation
\[
 \bar\phi_{\bar y}=r^{-4}\phi_{yy}=-\la\phi,\quad
 \bar\phi_{\bar y\bar y}=-\la r^{-4}\phi_y=-\la\bar r^4\bar\phi.
\]
This transformation is equivalent to the Darboux transformation
(\ref{DT1:uf}) generated by the function $\psi^{(0)}=r$. Indeed,
\[
 \bar\psi=\bar r\bar\phi=\phi_y/r=r\phi_x=r(\psi/r)_x=\psi_x-r_x\psi/r.
\]
Notice that the formulae (\ref{DT2}) correspond to the choice
$\psi^{(0)}=(y+c)r=r\int r^{-2}dx+cr$.

For the second transformation, one has
\begin{gather*}
 \ti\phi_{\ti y}=\phi_y\phi^{(\a)}-\phi\phi^{(\a)}_y,\\
 \ti\phi_{\ti y\ti y}
  =(\phi_{yy}\phi^{(\a)}-\phi\phi^{(\a)}_{yy})(\phi^{(\a)})^2
  =(\a-\la)r^4\phi(\phi^{(\a)})^3=(\a-\la)\ti r^4\ti\phi.
\end{gather*}
The change (\ref{LT0}) gives
\[
 \ti\psi=\ti r\ti\phi=\psi,\quad
 \ti u=\frac{\ti r_{xx}}{\ti r}=\frac{r_{xx}}{r}
 +\frac{1}{\phi^{(\a)}}\Bigl(\frac{2r_x}{r}\phi^{(\a)}_x+\phi^{(\a)}_{xx}\Bigr)
 =u-\a,
\]
that is, this transformation is equivalent to the shift
\[
 \ti\psi=\psi,\quad \ti u=u-\a,\quad \ti\la=\la-\a
\]
in the Schr\"odinger equation.
\end{proof}

\section{Dressing chains}\label{sec:dress}

When studying the iterations of the Darboux transformation (\ref{DT2}) it is
convenient to introduce an additional scaling by setting $\hat
r_n=-\g_nr_{n+1}$, $\hat y_n=\g^{-2}_ny_{n+1}$. Introduce the parameter $x$
accordingly to equation $dx=r^2_ndy_n$ and eliminate $p_n$ from relations
$p_n=-\g_nr_{n+1}r_n$, $p_{n,y_n}+p^2_n=-\a_nr^4_n$, then the sequence of
equations $(r_{n+1}r_n)_x=\g_nr^2_{n+1}+\a_n/\g_nr^2_n$ appears, which looks
more symmetric under the choice
\[
 -\g^2_n=\a_n.
\]
Accepting this we lose the Darboux transformation corresponding to $\a=0$.
However, it is clear that this transformation actually differs from the
other ones and it should be considered separately. This will be done in
\hyperref[sec:CH]{Section \ref*{sec:CH}}. So, we come to equations
\begin{equation}\label{dress:r}
 (r_{n+1}r_n)_x=\g_n(r^2_{n+1}-r^2_n),\quad y_{n,x}=r^{-2}_n.
\end{equation}
The change $r_n=e^{g_n}$ brings the first one to the form
\begin{equation}\label{dress:g}
 g_{n+1,x}+g_{n,x}=2\g_n\sinh(g_{n+1}-g_n)
\end{equation}
which defines the $x$-part of the B\"acklund transformation for the pot-mKdV
and $\sinh$-Gordon equations (see eg. \cite{Lamb80})
\[
 g_t=g_{xxx}-2g^3_x,\quad g_{x\tau}=\sinh2g.
\]
We will call the differential-difference equations of such type the {\em
dressing chains}. This terminology was suggested in the papers
\cite{Shabat_Yamilov91,Shabat92,Veselov_Shabat} in connection with the
equations
\begin{equation}\label{dress:f}
 f_{n+1,x}+f_{n,x}=f^2_n-f^2_{n+1}+\a_n-\a_{n+1}
\end{equation}
describing the iterations of Darboux transformation (\ref{DT1:uf}). The
transformations (\ref{DT1:uf}) and (\ref{DT2}) were shown in the previous
Section to be conjugated by Liouville transformation (\ref{LT0}). This
suggests that some relation exists between the chains (\ref{dress:r}) and
(\ref{dress:f}). In order to find it we will use the well-known
commutativity property of Darboux transformations, or the nonlinear
superposition principle (see eg. \cite{Lamb80}). This property is most
conveniently expressed in terms of the pre-potential introduced by the
formulae
\[
  2v_{n,x}=u_n,\quad v_n-v_{n+1}=f_n
\]
which bring to the following form of the dressing chain (\ref{dress:f}):
\begin{equation}\label{dress:v}
 v_{n+1,x}+v_{n,x}=(v_{n+1}-v_n)^2+\a_n.
\end{equation}

\begin{statement}\label{st:NSP}
Let $v$, $v^{(\a)}$ and $v^{(\b)}$ be related by Darboux
transformations
\[
 v^{(\a)}_x+v_x=(v^{(\a)}-v)^2+\a,\quad
 v^{(\b)}_x+v_x=(v^{(\b)}-v)^2+\b.
\]
Then the quantity $v^{(\a,\b)}=v-\dfrac{\a-\b}{v^{(\a)}-v^{(\b)}}$ is
related to $v^{(\a)}$ and $v^{(\b)}$ by Darboux transformations
\[
 v^{(\a,\b)}_x+v^{(\b)}_x=(v^{(\a,\b)}-v^{(\b)})^2+\a,\quad
 v^{(\a,\b)}_x+v^{(\a)}_x=(v^{(\a,\b)}-v^{(\a)})^2+\b.
\]
\end{statement}

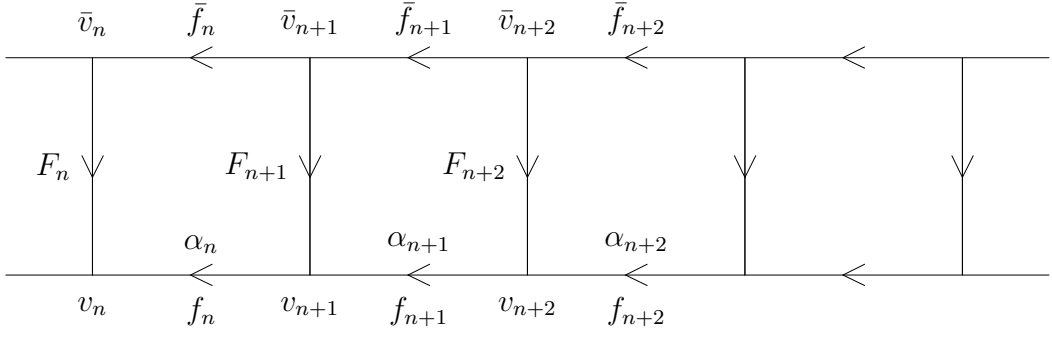
\begin{figure}[tb]
\begin{center}\setlength{\unitlength}{0.07em}
\begin{picture}(500,150)(-50,-30)
 \multiput(0,0)(0,100){2}{\path(-40,0)(440,0)
   \multiput(45,0)(100,0){4}{\path(10,5)(0,0)(10,-5)}}
 \multiput(0,0)(100,0){5}{\path(0,0)(0,100)}
 \multiput(0,0)(100,0){5}{\path(0,0)(0,100)\path(-5,55)(0,45)(5,55)}
 \multiputlist(0,-10)(50,0)[ct]{$v_n$,$f_n$,$v_{n+1}$,$f_{n+1}$,$v_{n+2}$,$f_{n+2}$}
 \multiputlist(50,10)(100,0)[cb]{$\a_n$,$\a_{n+1}$,$\a_{n+2}$}
 \multiputlist(0,110)(50,0)[cb]{$\bar v_n$,$\bar f_n$,$\bar v_{n+1}$,
      $\bar f_{n+1}$,$\bar v_{n+2}$,$\bar f_{n+2}$}
 \multiputlist(-10,50)(100,0)[rc]{$F_n$,$F_{n+1}$,$F_{n+2}$}
\end{picture}
\caption{two copies of the dressing chain}
\label{fig:kolbasa}
\end{center}
\end{figure}

Now, consider two copies of the dressing chain (\ref{dress:v}) with respect
to the variables $v_n$ and $\bar v_n$ related by Darboux transform with zero
parameter:
\begin{equation}\label{vv}
 v_{n,x}+\bar v_{n,x}=(v_n-\bar v_n)^2.
\end{equation}
The consistency of these equations with both copies of the chain is provided
by the \hyperref[st:NSP]{Statement \ref*{st:NSP}}, and moreover the
following relations are fulfilled
\begin{equation}\label{vvvv}
 (v_{n+1}-\bar v_n)(v_n-\bar v_{n+1})=\a_n.
\end{equation}
The differences $f_n$, $\bar f_n$ are identified with the oriented
horizontal edges of the lattice shown on the
\hyperref[fig:kolbasa]{fig.~\ref*{fig:kolbasa}}. It is not difficult to
prove that the differences $F_n=v_n-\bar v_n$ corresponding to the vertical
edges satisfy the chain
\begin{equation}\label{dress:F}
 F_{n+1,x}+F_{n,x}=(F_{n+1}-F_n)\sqrt{(F_{n+1}+F_n)^2-4\a_n}.
\end{equation}
Indeed,
\begin{align*}
 (F_{n+1}+F_n)_x &= (v_{n+1}+v_n-\bar v_{n+1}-\bar v_n)_x
  = (v_{n+1}-v_n)^2-(\bar v_{n+1}-\bar v_n)^2 \\
 &= (v_{n+1}-v_n-\bar v_{n+1}+\bar v_n)(v_{n+1}-v_n+\bar v_{n+1}-\bar v_n) \\
 &= (F_{n+1}-F_n)\sqrt{(v_{n+1}+v_n-\bar v_{n+1}-\bar v_n)^2
     -4(v_{n+1}-\bar v_n)(v_n-\bar v_{n+1})} \\
 &\overset{(\ref{vvvv})}{=} (F_{n+1}-F_n)\sqrt{(F_{n+1}+F_n)^2-4\a_n}.
\end{align*}
The chains (\ref{dress:F}) and (\ref{dress:r}) are related by the
substitution $F_n=r_{n,x}/r_n$. Remind, that the operator $D_x-r_x/r$
defines the special Darboux transformation at $\a=0$ (see the proof of the
\hyperref[st:invshift]{Statement \ref*{st:invshift}}).

Coming back to the chain (\ref{dress:f}), notice that the relation
(\ref{vvvv}) is equivalent to the quadratic equation with respect to $f_n$.
Its solution provides the substitutions into the chain (\ref{dress:f}) and
its copy for the variables $\bar f_n$.

\begin{statement}\label{st:ffr}
The general solutions of the chains (\ref{dress:f}), (\ref{dress:F}) and
(\ref{dress:r}) are related by the substitutions
\begin{gather}
\nonumber
 F_n=\frac{r_{n,x}}{r_n},\quad
 s_n:=\sqrt{(F_{n+1}+F_n)^2-4\a_n}
     =\g_n\Bigl(\frac{r_{n+1}}{r_n}+\frac{r_n}{r_{n+1}}\Bigr),\quad
 \a_n=-\g^2_n,    \\
\nonumber
 2f_n=F_n-F_{n+1}-s_n,\quad 2\bar f_n=F_{n+1}-F_n-s_n,\\
\label{fr}
 f_n=\frac{r_{n,x}-\g_nr_{n+1}}{r_n},\quad
 \bar f_n=-\frac{r_{n,x}}{r_n}-\frac{\g_nr_n}{r_{n+1}}.
\end{gather}
\end{statement}

Till now we have not paid attention to the variables $y_n$. It turns out
that the dressing chain can be rewritten in terms of these variables as
well. One obtains, after dividing (\ref{dress:r}) by $r^2_{n+1}r^2_n$ and
integrating (the integration constants are unessential here and are
eliminated by the shift $y_n\to y_n+c_n$), the relation
\begin{equation}\label{yr}
 \frac1{\g_nr_{n+1}r_n}=y_{n+1}-y_n
\end{equation}
which means that all $y_n$ can be recovered by the single quadrature. These
equations can be rewritten also in the form
\begin{equation}\label{dress:y}
 y_{n+1,x}y_{n,x}=\g^2_n(y_{n+1}-y_n)^2.
\end{equation}
Finally, notice that the change $\bar r_n=1/r_n$ arising from the
transformation (\ref{inv}) is equivalent to the reversion of the vertical
arrows on the \hyperref[fig:kolbasa]{fig.~\ref*{fig:kolbasa}} and it leads
to the chains
\[
 (\bar r_{n+1}\bar r_n)_x=\g_n(\bar r^2_{n+1}-\bar r^2_n),\quad
 \bar y_{n+1,x}\bar y_{n,x}=\g^2_n(\bar y_{n+1}-\bar y_n)^2,\quad
 \bar y_{n,x}=\bar r^{-2}_n.
\]
The variables $\bar y_n$ satisfy the recurrent relation analogous to
(\ref{yr}):
\begin{equation}\label{byr}
 \bar y_{n+1}-\bar y_n=r_{n+1}r_n/\g_n.
\end{equation}

\section{Dym equation}\label{sec:Dym}

The dressing chains introduced in the previous Section belong to the rather
important class of the B\"acklund transformations of the general form
\begin{equation}\label{ux}
  u_{n+1,x}=b(u_{n,x},u_n,u_{n+1},\a_n)
\end{equation}
which correspond to the KdV type equations
\begin{equation}\label{utA}
  u_t=A(u_{xxx},u_{xx},u_x,u,x,\a).
\end{equation}
For example, the chains (\ref{dress:f}) and (\ref{dress:F}) define the
$x$-parts of the B\"acklund transformations for the modified KdV equations
\[
 f_t=f_{xxx}-6(f^2+\a)f_x,\quad F_t=F_{xxx}-6F^2F_x,
\]
while the chains (\ref{dress:y}) and (\ref{dress:v}) correspond to the
Schwarz-KdV and pot-KdV equations
\[
 y_t=y_{xxx}-\frac{3y^2_{xx}}{2y_x},\quad v_t=v_{xxx}-6v^2_x.
\]
As we have already mentioned, the chain (\ref{dress:g}) corresponds to the
pot-mKdV equation or, in the $r$ variables,
\begin{equation}\label{rt}
 r_t=r_{xxx}-\frac{3r_{xx}r_x}{r}.
\end{equation}
Although the list can be easily expanded, it does not fall outside the
subclass of equations of the form
\begin{equation}\label{uta}
 u_t=u_{xxx}+a(u_{xx},u_x,u,x,\a).
\end{equation}
Indeed, the compatibility of the equations (\ref{ux}), (\ref{utA}) means
that the identity holds
\[
 D_x(A[n+1])=b_{u_{n,x}}D_x(A[n])+b_{u_n}A[n]+b_{u_{n+1}}A[n+1]
\]
where the derivatives of $u_{n+1}$ are eliminated in virtue of (\ref{ux}).
Equating the coefficients at $u_{n,xxxx}$ gives the relation
$A_{u_{xxx}}[n+1]=A_{u_{xxx}}[n]$. The quantity $A_{u_{xxx}}^{-1/3}$ is
called the separant of equation (\ref{utA}). If it depends on $x$ only then
the equation can be brought to the form (\ref{uta}). Otherwise the chain
(\ref{ux}) does not lead out of the finite-parametric family of solutions of
the ordinary differential equation of the form
$A_{u_{xxx}}=A_{u_{xxx}}[0]=c(x)$. It is clear that such type of B\"acklund
transformation must be considered as defective.

Nevertheless, a chain of the form (\ref{ux}) may define a B\"acklund
transformation for an equation with variable separant if it can be extended
by some equation for the independent variable, which we will denote as $y_n$
from now on. The variable $x$ now will play the role of an auxiliary
parameter. The chain (\ref{dress:r}) gives the example of such an extension,
corresponding to the Dym equation for the variable $w=r^{-2}$:
\begin{equation}\label{Dym}
 w_t=w^3w_{yyy}.
\end{equation}
Indeed, this equation is related to (\ref{rt}) by the following composition
of the hodograph transform and two differential substitutions like
introducing of the potential:
\[
 \begin{array}{ccc}
  x_y=\dfrac{1}{w},\quad x_t=\dfrac{1}{2}w^2_y-ww_{yy} &&
  y_x=\dfrac{1}{r^2},\quad y_t=-\dfrac{2r_{xx}}{r^3}\\[5mm]
  \Downarrow && \Downarrow \\[2mm]
  x_t=\dfrac{x_{yyy}}{x^3_y}-\dfrac{3x^2_{yy}}{2x^4_y}
  &\quad\Leftrightarrow\quad &
  y_t=y_{xxx}-\dfrac{3y^2_{xx}}{2y_x}
 \end{array}
\]
(such type of changes can be written a little bit more compactly as a
reciprocal transformation). The relation $w=r^{-2}$ is just a consequence of
the identity $y_xx_y=1$.

\begin{remark}\label{rem:ext}
An extension of the chain may be not unique. Indeed, equation (\ref{rt})
admits a more general potential:
\[
 y_x=ar^{-2}+br^2,\quad y_t=-2ar_{xx}r^{-3}+2b(rr_{xx}-2r^2_x).
\]
It can be used in order to preserve the real structure when $r\sim e^{ig}$,
that is in the case of sine-Gordon equation and mKdV equation with the plus
sign before the nonlinear term. In particular, the chain
\[
 g_{n+1,x}+g_{n,x}=2\g_n\sin(g_{n+1}-g_n)
\]
can be extended by setting $y_{n,x}=c+\sin2g_n$, where the constant $c\ge1$
is added in order to provide the one-to-one correspondence $x\leftrightarrow
y$ on the whole axis. This leads to the following sequence of the
transformations:
\[
 \begin{array}{ccc}
  w_t=w^3w_{yyy}-\dfrac32w^2D_y\Bigl(\dfrac{w^2_y(1-c^2+cw)}{(w-c)^2-1}\Bigr) &&
  g_t=g_{xxx}+2g^3_x \\[4mm]
  \Uparrow && \Uparrow\\[3mm]
  x_y=\dfrac{1}{w},\quad
  x_t=\dfrac{1}{2}w^2_y-ww_{yy}+\dfrac{3w^2_y(1-c^2+cw)}{2((w-c)^2-1)} &&
  \begin{array}{l}y_x=c+\sin2g,\\ y_t=2g_{xx}\cos2g+2g^2_x\sin2g\end{array}\\[5mm]
  \Downarrow && \Downarrow \\[2mm]
  x_t=\dfrac1{x^3_y}\left(x_{yyy}
   -\dfrac{3x^2_{yy}(1-3cx_y+2(c^2-1)x^2_y)}{2x_y((1-cx_y)^2-x^2_y)}\right)
  &\ \Leftrightarrow\ &
  y_t=y_{xxx}-\dfrac{3(y_x-c)y^2_{xx}}{2((y_x-c)^2-1)}
 \end{array}
\]
By construction, the equation for $w$ must support the breather type
solutions. However, it is not clear, if it possesses some applications, so
we will not discuss it further on.
\end{remark}

The construction of the extended dressing chain can be based on the zero
curvature representation
\[
 \Phi_y=M\Phi,\quad \Phi_t=N\Phi \quad\Rightarrow\quad M_t=N_y+[N,M].
\]
Since, in the case of the variable separant, the B\"acklund transformation
involves the independent variable, hence it is natural to replace $\pd_y$ by
differentiation with respect to some parameter $x$ independent on $n$,
assuming
\[
 \pd_x=\rho_n\pd_{y_n},\quad \pd_T=\pd_{t_n}+\sigma_n\pd_{y_n},\quad
 \rho_n=y_{n,x},\quad \sigma_n=y_{n,T}.
\]
Then the compatibility condition of the auxiliary linear problems
\[
 \Phi_{n,x}=\rho_nM_n\Phi_n,\quad \Phi_{n+1}=L_n\Phi_n
\]
defines the extended dressing chain
\begin{equation}\label{Lx}
 L_{n,x}=\rho_{n+1}M_{n+1}L_n-\rho_nL_nM_n,\quad y_{n,x}=\rho_n.
\end{equation}
If the matrix $M$ is given then this equation allows to find constructively
both the factor $\rho_n$ and the matrix $L_n$. Analogously, the $t$-part of
the B\"acklund transformation is derived from the consistency with the
linear problem
\[
 \Phi_{n,T}=(N_n+\sigma_nM_n)\Phi_n,\quad y_{n,T}=\sigma_n.
\]
For example, the Dym equation defines the isospectral deformation of the
acoustic problem (\ref{Dy:r}):
\[
 \phi_{yy}=-\la w^{-2}\phi,\quad \phi_t=2\la w_y\phi-4\la w\phi_y.
\]
In the matrix form,
\[
 \Phi=\begin{pmatrix} \phi \\ \phi_y \end{pmatrix},\quad
 M=\begin{pmatrix} 0 & 1 \\ -\la w^{-2} & 0 \end{pmatrix},\quad
 N=2\la\begin{pmatrix} w_y & -2w \\ 2\la w^{-1}+w_{yy} & -w_y \end{pmatrix}.
\]
Extension of the Darboux transformation (\ref{DT2}) on $\hat\phi_{\hat y}$
brings to the matrix
\[
 L_n=\begin{pmatrix} \g_n & (r_{n+1}r_n)^{-1}\\
          -\la r_{n+1}r_n & \g_n \end{pmatrix},
\]
and the substitution into (\ref{Lx}) yields the equations of the chain along
with the constraint
\[
 \rho_n=w_n=r^{-2}_n.
\]
Conversely, it is easy to check that if we assume $\deg_\la L_n=\deg_\la\det
L_n=1$ then both this constraint and the matrix $L_n$ itself are found
uniquely from (\ref{Lx}).

\begin{remark}
The hypothesis exists that all integrable equations of the form (\ref{utA})
can be brought via the differential substitutions and contact or point
transformations to the form with constant separant. If this is true then the
B\"acklund transformations for equations (\ref{utA}) may be obtained from
the B\"acklund transformations for equations (\ref{uta}) by means of
suitable extension, like in the above examples. The integrable equations of
the form (\ref{uta}) are very well studied and it is known that any of them
can be reduced (via the transformations not involving $x$) either to KdV or
Krichever-Novikov or linear equation. Correspondingly, all dressing chains
can be reduced finally to few basic ones. On the contemporary state of the
problem of classification of the equations (\ref{utA}) see \cite{Heredero}.
\end{remark}

\section{Camassa-Holm equation}\label{sec:CH}

The Camassa-Holm equation \cite{Fokas_Fuchssteiner,Camassa_Holm}
\begin{equation}\label{CH}
 4h_t-h_{zzt}+2\eps h_z=hh_{zzz}+2h_zh_{zz}-12hh_z
\end{equation}
appears as the compatibility condition for the linear problems
\[
 \chi_{zz}=(1-\la R^4(z))\chi,\quad
 \chi_t=\frac{h_z}{2}\chi+\Bigl(\frac{1}{2\la}-h\Bigr)\chi_z,\quad
 R^4:=h_{zz}-4h-\eps.
\]
The dressing chain for this equation, with respect to the variables $z,R$,
may be obtained simply by applying the Liouville transformation
(\ref{LT:rR}) to the chain (\ref{dress:y}):
\begin{equation}\label{dress:z}
 z_{n+1,x}z_{n,x}=\g^2_n\sinh^2(z_{n+1}-z_n),\quad z_{n,x}=R^{-2}_n.
\end{equation}
The equation (\ref{CH}) is equivalent to the conservation law
$(R^2)_t+(R^2h)_z=0$ which allows to apply the reciprocal transformation
\[
 dx=R^2dz-R^2hdt \quad\Rightarrow\quad
 z_x=R^{-2},\quad z_t=h.
\]
Eliminating $R$ and $h$ from the equality $R^4=(z^{-1}_xD_x)^2(h)-4h-\eps$
yields the equation
\begin{equation}\label{zxxt}
 z_{xxt}z_x-z_{xt}z_{xx}=(4z_t+\eps)z^3_x+z_x
\end{equation}
which is one of the equivalent forms of the so called associated
Camassa-Holm equation \cite{Schiff,Hone,Ivanov}. The first equation of the
extended chain (\ref{dress:z}) defines the $x$-part of the B\"acklund
transformation for the equation (\ref{zxxt}).

Along with this transformation, equation (\ref{zxxt}) admits one more
B\"acklund transformation which does not contain the parameter and may be
considered as a limiting case. The interesting fact is that the
corresponding $t$-part can be written as a Volterra type lattice. In order
to avoid confusion we will denote the iterations of this B\"acklund
transformation by superscript.

\begin{statement}
The following two lattices commute:
\begin{equation}\label{zmxt}
 z^m_xz^{m+1}_x=e^{2z^m-2z^{m+1}},\quad
 -8z^m_t=2\eps+e^{2z^{m+1}-2z^m}+e^{2z^m-2z^{m-1}}.
\end{equation}
The variables $z^m$ solve equation (\ref{zxxt}) in virtue of these lattices.
\end{statement}

Several equivalent equations can be written down. Coming back to the
Schr\"odinger gauge, assume $\psi=R\chi$ then
\[
 \psi_{xx}=(u-\la)\psi,\quad 2\la\psi_t=R^2\psi_x-RR_x\psi,\quad
 u=\frac{R_{xx}}{R}+\frac1{R^4},\quad u_t=-2RR_x.
\]
The variable $v$ is introduced accordingly to the relations $u=2v_x$,
$R^2=-2v_t$ and elimination of $R$ yields the equation
\[
 2v_tv_{xxt}-v^2_{xt}-8v_xv^2_t+1=0
\]
which is represented by the consistent pair of the lattices
\begin{equation}\label{vmxt}
 v^m_x+v^{m+1}_x=(v^m-v^{m+1})^2,\quad
 v^m_t=(v^{m+1}-v^{m-1})^{-1}.
\end{equation}
The differences $F^m=v^m-v^{m+1}$ correspond to the pair
\[
 F^m_x+F^{m+1}_x=(F^m)^2-(F^{m+1})^2,\quad
 F^m_t=(F^{m+1}+F^m)^{-1}-(F^m+F^{m-1})^{-1}.
\]
Its consistency (and few other equivalent forms) was stated in the papers
\cite{Shabat_Yamilov91,Yamilov90}, although the associated equation
\[
 FF_{xxt}-F_xF_{xt}-4F^3F_t+2F_x=0
\]
was not explicitly presented.

Obviously, the first of the chains (\ref{vmxt}) governs the iterations of
the Darboux transformations with zero parameter (\ref{vv}). Therefore, this
B\"acklund transformation is interpreted as the reproduction of the
\hyperref[fig:kolbasa]{fig.~\ref*{fig:kolbasa}} in vertical direction.
It is easy to check that the totally discrete equation implied by
(\ref{vvvv})
\[
 (v^m_{n+1}-v^{m+1}_n)(v^m_n-v^{m+1}_{n+1})=\a_n
\]
is consistent with the dynamics on $t$.

\section{Formulae with Wronskians}

Let the wave functions $r_1=\psi^{(0)}_1$ and $\psi^{(\a_n)}_1$ are known
for the Schr\"odinger operator $-D^2_x+u_1$, corresponding to the pairwise
different values of the spectral parameter (in the case of the multiple
parameters all formulae below remain valid, but one should to use the
associated functions in addition to the wave ones). The Darboux
transformation defined by the function $\psi^{(\a_1)}_1$ leads to the
potential $u_2$ with the wave functions
$\psi^{(\a_n)}_2=(D_x-f_1)(\psi^{(\a_n)}_1)$, $n\ne1$, where
$f_1=\psi^{(\a_1)}_{1,x}/\psi^{(\a_1)}_1$. Further on, application of the
transformations defined by the functions $\psi^{(\a_2)}_2$,
$\psi^{(\a_3)}_3$ etc, generates the sequence of the multiple Darboux
transformations. Their result is expressed in terms of the Wronskians of the
original functions $\psi^{(\a_n)}_1$ accordingly to the Crum formulae
\cite{Crum}
\begin{gather*}
 \psi^{(\la)}_{n+1}=\D_n(\psi^{(\la)}_1)/\D_n,\quad
 f_{n+1}=D_x\log(\D_{n+1}/\D_n),\\
 u_{n+1}=u_1-2D^2_x\log\D_n,\quad v_{n+1}=v_1-D_x\log\D_n, \\
 \D_0=1,\quad \D_n=\<\psi^{(\a_1)}_1,\dots,\psi^{(\a_n)}_1\>,\quad
 \D_n(g)=\<\psi^{(\a_1)}_1,\dots,\psi^{(\a_n)}_1,g\>,\\
 \<g_1,\dots,g_n\>:=\det(D^{k-1}_x(g_j))|^n_{j,k=1}.
\end{gather*}
The proof follows immediately from the definition of the Darboux
transformation (\ref{DT1:uf}) if one takes into account the identity
\[
 \<g_1,\dots,g_n\>=g_1\<A(g_2),\dots,A(g_n)\>,\quad A=D_x-g_{1,x}/g_1.
\]
The Crum formulae can be easily extended on the other variables introduced
in \hyperref[sec:dress]{Section~\ref*{sec:dress}}. Since $r_1$ is a wave
function at $\la=0$, hence
\[
 \bar v_n=v_1-D_x\log\D_{n-1}(r_1),\quad F_n=D_x\log(\D_{n-1}(r_1)/\D_{n-1}).
\]
Therefore $r_n=c_n\D_{n-1}(r_1)/\D_{n-1}$, and the multiplier can be found
from the formula for $f_n$ (\ref{fr}) rewritten in the form
\[
 \<\D_n,\D_{n-1}(r_1)\>=\g_n\frac{c_{n+1}}{c_n}\D_{n-1}\D_n(r_1).
\]
The Jacobi identity
\[
 \<\<g_1,\dots,g_n,g),\<g_1,\dots,g_n,h\>\>=\<g_1,\dots,g_n\>\<g_1,\dots,g_n,g,h\>,
\]
implies $c_{n+1}=c_n/\g_n=(\g_n\dots\g_1)^{-1}$. Finally, we come to the
following parametric representation of the potentials and wave functions of
the acoustic problem:
\begin{gather}
\label{Wr:yr}
 r_{n+1}=\frac{\D_n(r_1)}{\g_n\dots\g_1\D_n}, \quad
 \phi^{(\la)}_{n+1}=\frac{\D_n(\psi^{(\la)}_1)}{\D_n(r_1)},\quad
 y_{n+1}=\sum^n_{k=1}\frac{1}{\g_kr_{k+1}r_k}+\int r^{-2}_1(\xi)d\xi,\\
\label{Wr:byr}
 \bar r_{n+1}=\frac1{r_{n+1}}, \quad
 \bar\phi^{(\la)}_{n+1}=\frac{\D_n(r_1,\psi^{(\la)}_1)}{\D_n},\quad
 \bar y_{n+1}=\sum^n_{k=1}\frac{r_{k+1}r_k}{\g_k}+\int r^2_1(\xi)d\xi.
\end{gather}

\paragraph{Example 1.}
Let $u_1=c^2$, $\a_n=c^2-\ka^2_n$ where $0<\ka_1<\dots<\ka_{N-1}<\ka_N=c$.
The wave functions are assumed to be
\[
 \psi^{(\a_n)}_1=\left\{\begin{array}{ll}
  \cosh(\ka_nx+\d_n), & n=2k-1 \\
  \sinh(\ka_nx+\d_n), & n=2k
  \end{array}\right.,\quad
 r_1=\psi^{(\a_N)}_1=\psi^{(0)}_1.
\]
Then the determinants $\D_n$ do not vanish on the real axis and
asymptotically $\D_n\sim e^{(\ka_1+\dots+\ka_n)|x|}$, $x\to\infty$. Since
$\D_{N-1}(r_1)=\D_N$ in virtue of the definition of $r_1$, hence the
potentials $u_N=c^2-2D^2_x\log\D_{N-1}$ and $\bar u_N=c^2-2D^2_x\log\D_N$
are the usual multisoliton potentials of the Schr\"odinger operator raised
on $c^2$.

\begin{figure}[tb]
\center{\begin{tabular}{ccc}
 \includegraphics[width=50mm]{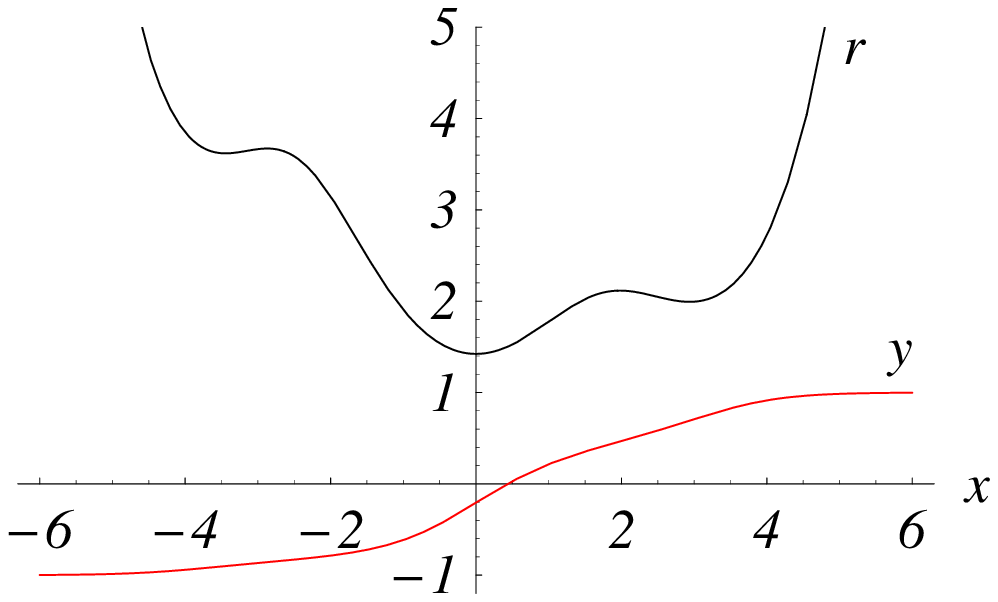}&
 \includegraphics[width=50mm]{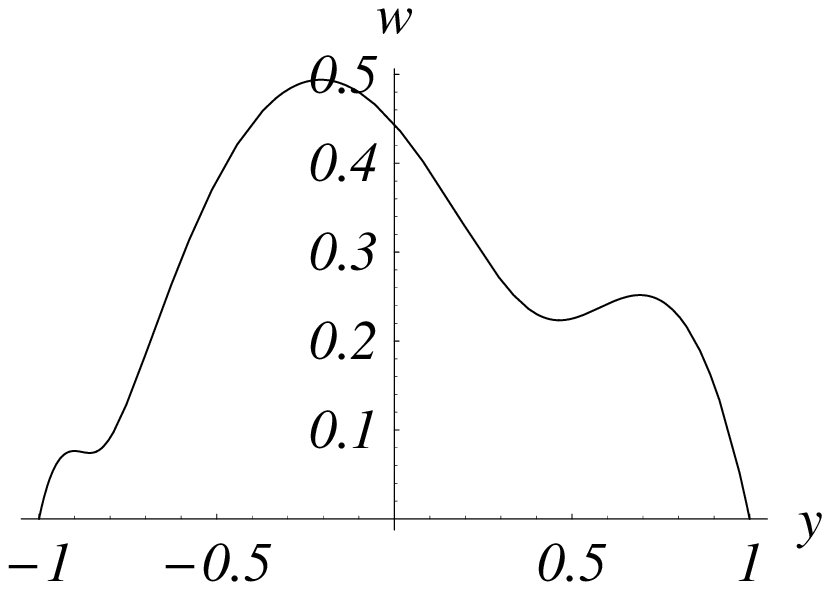}&
 \includegraphics[width=50mm]{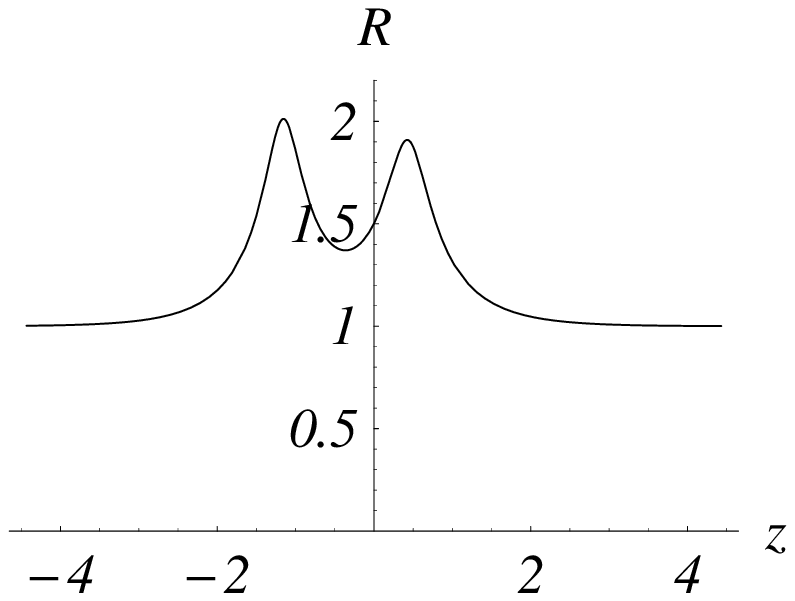}\\
 $(a)$ & $(b)$ & $(c)$\\
 \includegraphics[width=50mm]{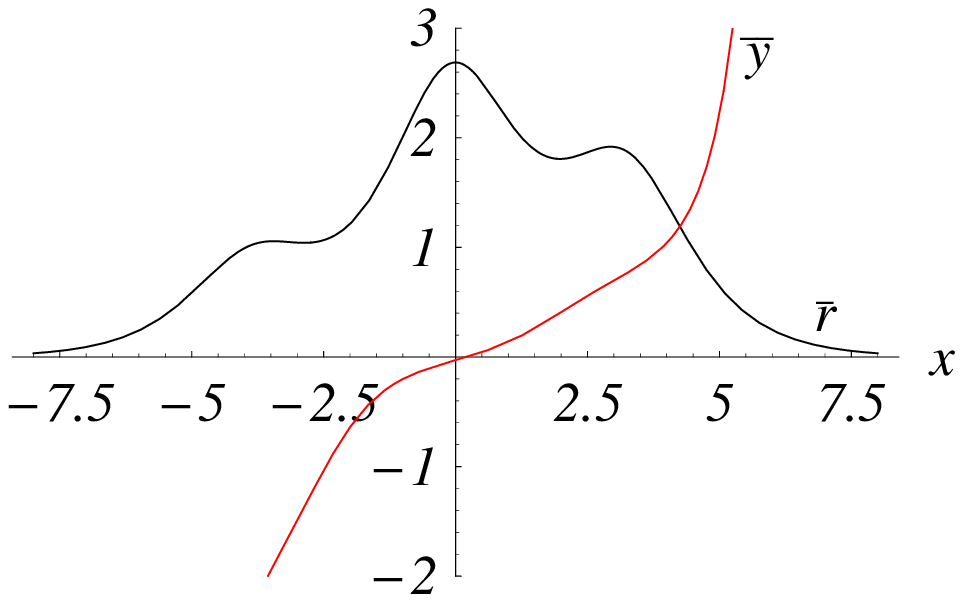}&
 \includegraphics[width=50mm]{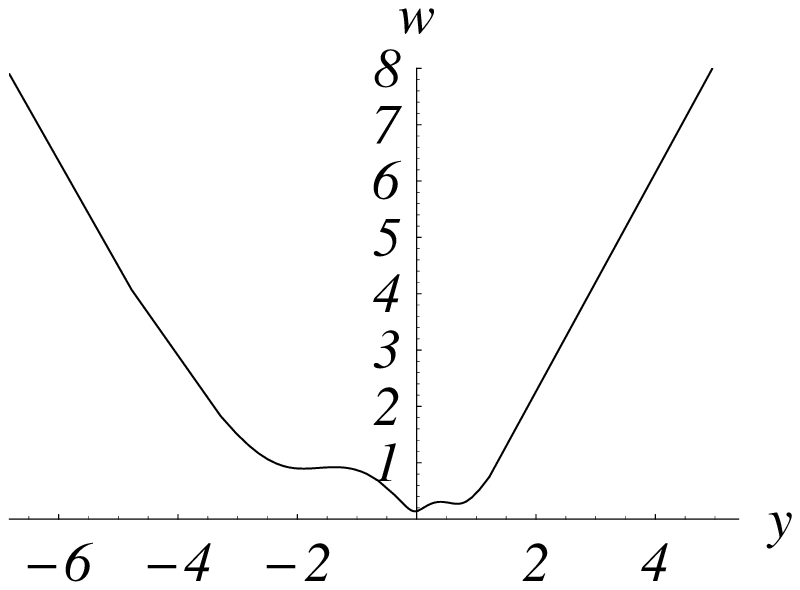}&
 \includegraphics[width=50mm]{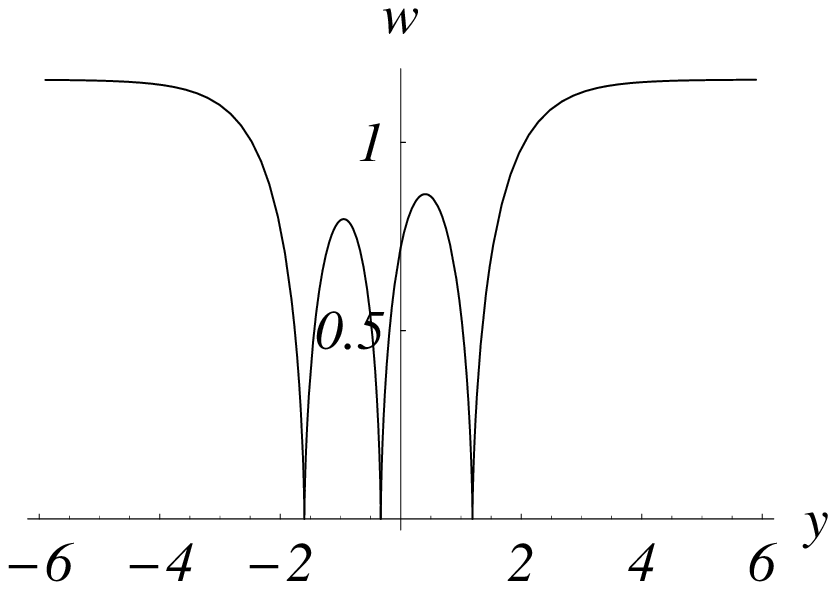}\\
 $(d)$ & $(e)$ & $(f)$
\end{tabular}}
\caption{($a$)--($e$) 3-soliton potential in various coordinates;}
 ($f$) irregular potential. \label{fig:sol3}
\end{figure}

The potentials of the acoustic problem, as functions on $x$, have the
asymptotic behavior $r_N\sim\cosh cx$, $\bar r_N\sim\sech cx$. They are
shown on the \hyperref[fig:sol3]{fig.~\ref*{fig:sol3}a,d}, together with the
corresponding arguments $y_N$, $\bar y_N$. Both functions $y_N$, $\bar y_N$
are monotonically increasing, but $y_N$ is bounded while $\bar y_N$ grows at
infinity as $\sinh2cx$. Consider these two cases separately.

1) Choose the value of the antiderivative in (\ref{Wr:yr}) equal to
$c^{-1}\tanh(cx+\d_N)$ if $N=2k-1$ and $c^{-1}\coth(cx+\d_N)$ if $N=2k$.
Then it is easy to prove that the scaled variable $y=cy_N$ changes in the
limits from $-1$ till $1$ and the graph $w(y)=cr^{-2}_N$ looks like a finite
cap (\hyperref[fig:sol3]{fig.~\ref*{fig:sol3}b}). The phase dependence on
$t$ of the form
\begin{equation}\label{dn}
 \d_n=\ka_n(4\ka^2_n-6c^2)t+\ti\d_n,\quad \ti\d_n=\const
\end{equation}
brings to the solution of the Dym equation (\ref{Dym}) on the interval
$[-1,1]$ with zero boundary conditions.

Next, application of the Liouville transformation (\ref{LT:rR}) yields the
function $R(z)$ in the parametric form
(\hyperref[fig:sol3]{fig.~\ref*{fig:sol3}c})
\[
 R=r_N(x)\sqrt{1-y^2(x)},\quad z=\frac12\log\frac{1+y(x)}{1-y(x)}.
\]
Notice that one of the solitons is ``absorbed'' by the transformation. One
can prove that the phase dependence on $t$ corresponding to Camassa-Holm
equation is of the form
\[
 \d_n=\frac{\ka_nt}{2c(c^2-\ka^2_n)}+\ti\d_n,\quad n=1,\dots,N-1,\quad
 \d_N=-\frac{(1+\eps c^2)t}{4c^2}+\ti\d_N.
\]

2) The function $w(y)=r^2_N$, $y=\bar y_N$ has asymptotically linear
behavior (\hyperref[fig:sol3]{fig.~\ref*{fig:sol3}e}). It defines the
solution of Dym equation if the phases $\d_n$ depend on $t$ accordingly to
(\ref{dn}) and the value of the antiderivative is
\[
 \bar y_1=\frac1{4c}\sinh2\d_N+(-1)^{N-1}\Bigl(\frac{x}2+3c^2t\Bigr)+\const.
\]

\paragraph{Example 2.}
Let $u_1=0$, $\a_n=-\g^2_n$, where $\g_N=0<\g_1<\dots<\g_{N-1}$,
\[
 \psi^{(\a_n)}_1=\left\{\begin{array}{ll}
  \cosh(\g_nx+4\g^3_nt+\d_n), & n=2k-1 \\
  \sinh(\g_nx+4\g^3_nt+\d_n), & n=2k
  \end{array}\right.,\quad
 r_1=\psi^{(0)}_1=1.
\]
Then the function $r_N(x)=\D_{N-1}(1)/\D_{N-1}$ has $N-1$ zeroes. Take $\bar
y_1=x$, then the formulae $w(y)=r^2_N$, $y=\bar y_N$ define a solution of
Dym equation with $N-1$ singularities
(\hyperref[fig:sol3]{fig.~\ref*{fig:sol3}f}).

\section{Concluding remarks}

The main conclusion of the paper can be formulated as follows: in the case
of an equation with variable separant it is convenient to represent the
chain of B\"acklund transformations in the parametric form as a
two-component chain for dependent and independent variables. We have
illustrated this by the example of Dym equation where the dressing chain is
an extension of the dressing chain for mKdV equation. It is generated by
Darboux transformations for the Schr\"odinger operator in the acoustic
gauge, in contrast to the dressing chain for the KdV equation which
corresponds to the standard gauge. Both versions merge on the totally
discrete level. Camassa-Holm equation finds its place in this picture as
well.

Many problems remain beyond the scope of our paper, for example the
ultrashort pulse equation \cite{Schafer_Wayne} related to the sine-Gordon
equation. Possibly, the \hyperref[rem:ext]{Remark \ref*{rem:ext}} suggests
the way of construction of the B\"acklund transformations in this case as
well, but the Zakharov-Shabat spectral problem is more likely to be
convenient to start from. The two-component analogs of Dym and Camassa-Holm
equations which have drawn the attention only recently (see eg.~\cite{AGZ})
are also associated to this spectral problem. Comparing to the Schr\"odinger
operator, this case admits more diversity in the Darboux transformations
which generate, in particular, the lattices of the Toda and relativistic
Toda lattice type.

The recent Degasperis-Procesi equation
\cite{Degasperis_Procesi,Degasperis_Hone_Holm} is of some interest as well.
This equation is very close to Camassa-Holm one, but it is associated with
the Kaup-Kupershmidt spectral problem of the third order. This makes the
realization of the general scheme more difficult since the Darboux
transformation for this problem is rather complicated, see
eg.~\cite{Adler_Marikhin_Shabat}.

The construction of the extended chains of Darboux and Laplace
transformations for $2+1$-dimensional equations with variable separant seems
to be straightforward. As a typical example one can take the generalized Dym
equation studied in the papers
\cite{Dubrovsky_Konopelchenko,Dmitrieva_Khlabystova95b}.

\bigskip

The authors thank V.V.~Soklov and M.V.~Pavlov for useful discussions. This
work was supported by the RFBR grant \# 04-01-00403.


\end{document}